\documentstyle[12pt]{article}
\def\normal{{\bigcirc \!\hskip -5pt n}} 

\def\0barra{{\rm O} \!\hskip -3.7pt {\rm l} } 

\def\1barra{1\! \hskip -1.1pt {\rm l}} 

\title{Grassmann Algebra  and  Fermions at Finite Temperature }

\vspace{-0.5cm}

\author{ I.C. Charret$^{(1)}$,  E.V. Corr\^ea Silva$^{(2)}$,
       S.M. de Souza$^{(1)}$, \\
   O. Rojas Santos$^{(3)}$ \hspace{3pt} and \hspace{3pt} 
           M.T. Thomaz$^{(3)}$ \thanks{Corresponding author: 
Dr. Maria Teresa Thomaz;
 R. Domingos S\'avio Nogueira Saad n.$\!\!^{\rm o}$ 120  apto 404, 
 Niter\'oi, R.J., 24210--310, BRAZIL
 --Phone: (21) 620-6735; Fax: (21) 620-3881;  {\it E-mail}: mtt@if.uff.br.
}  \\
\\
\baselineskip =10pt
{ \small \it $^{(1)}$ Departamento de Ci\^encias Exatas \vspace{-0.2cm}}\\
{ \small \it Universidade Federal de Lavras \vspace{-0.2cm}}\\
{\small \it Caixa Postal  37 \vspace{-0.2cm} }\\
{ \small \it CEP: 37200-000, Lavras, MG,  Brazil }\\
 { \small \it $^{(2)}${\small\it  Centro Brasileiro de 
	Pesquisas F\'\i sicas} \vspace{-0.2cm}	}\\
{\small \it R. Dr. Xavier Sigaud n$^{o}$ 150}  \vspace{-0.2cm}\\
{\small \it CEP: 22290-180, Rio de Janeiro, RJ, Brazil} \\ 
{ \small \it $^{(3)}$ Instituto de F\'\i sica \vspace{-0.2cm}}\\
{ \small \it Universidade Federal Fluminense  \vspace{-0.2cm}}\\
{\small \it Av. Gal. Milton Tavares de Souza s/n.$\!\!^\circ$, 
		 Niter\'oi,\vspace{-0.2cm} }\\
{ \small \it CEP: 24210-340, Rio de Janeiro, RJ, Brazil }   \\  }

\date{}

\begin{document}

\maketitle

\begin{abstract}

\baselineskip=14pt

For any $d$-dimensional self--interacting fermionic model, all
coefficients in the high--temperature expansion of its grand canonical
partition function can be put in terms of multivariable Grassmann
integrals. A new approach to calculate such coefficients, based on
direct exploitation of the grassmannian nature of fermionic operators,
is presented. We apply the method to the soluble Hatsugai-Kohmoto
model, reobtaining well-known results.

\end{abstract}

\vfill


\vspace{0.3cm}


\newpage

\baselineskip=18pt


\section{ Introduction}

A quantum system at thermal equilibrium can be completely described
provided that one knows its grand canonical partition function, which
can be expressed as a path integral. For bosonic systems, an
advantageous feature of the path integral approach is that of
employing commuting functions instead of non-commuting operators. For
fermionic systems, however, such an advantage is not obvious to hold,
as the integration variables are also non-commuting.

In 1980, Kubo\cite{kubo} used the path integral approach to calculate
the grand canonical partition function of the Hubbard model, using the
strong coupling limit and performing a perturbative expansion in the
hopping constant ($t$). Even though his result is valid for any
temperature, one does not have the exact coefficient of $\beta$ (
$\beta = \frac{1}{kT}$) of the high temperature expansion of the
partition function. Since then, improvements on the calculation of the
high temperature expansion up to order $(\beta t)^9$ for the Hubbard
model in two and three dimensions have been reported in the literature
\cite{henderson}.

Recently, Grandati {\it et al.}\cite{grandati} presented a method to
calculate the grand canonical partition function of self-interacting
fermions, by writing that function on a lattice and using the
properties of the Grassmann algebra to calculate its expansion in
powers of the coupling constant. They calculated the first two terms
for the bi-dimensional chiral Gross-Neveu model, obtaining an
analytical result; however, their approach is model-dependent. More
recently, Creutz\cite{creutz} used a numerical algorithm to calculate
the generating functional of a fermionic model, rewritten on a
lattice. He applied his algorithm to a unidimensional fermionic system
involving a thousand grassmannian variables. He pointed out that this
approach does not have the sign problems that generally hamper the
application of the Monte Carlo method to fermionic models.

We do not write the grand canonical partition function of a
self-interacting fermionic model on a lattice; instead, we present a
new method to obtain the coefficients of its high temperature
expansion in $d$-dimensions, where $d\geq 1$. But for the expansion in
$\beta$, this method does {\em not} involve any other perturbative
expansion (such as, say, in the couplig constant of the model).
In section 2 we present the method, an extension to the one used to
calculate the grand canonical partition function of the anharmonic
fermionic oscillator\cite{phys_A}, a quantum model in $d=0$
space-dimension. In reference \cite{BJP}, the properties of the
Grassmann algebra were  used to calculate the moments of
grassmannian gaussian integrals. This general result, together with
the diagonalization of matrices ${\bf A}^{\sigma \sigma}$ ($\sigma=
\uparrow, \downarrow$) --- matrices that appear when the trace of any
fermionic operator is expressed in terms of a multivariable Grassmann
integral (see eq.(\ref{11})) --- allows us to develop a general
approach to obtain analytical expressions for the coefficients of the
high temperature expansion of the grand canonical partition function
of any self-interacting fermioning model in $d$-dimensions, even in
the thermodynamical limit. In section 3 we apply the method to
Hatsugai-Kohmoto model, a simple toy model that was used by Hatsugai and
Kohmoto to explain the metal-insulator transition. The solution of
this model is very simple, and does not require all of the features
developed in section 2. In section 4 we present our conclusions and
future applications of the present approach. In appendix A, the
diagonalization of the matrices ${\bf A}^{\sigma \sigma}$, for
arbitrary lattice dimension and arbitrary number of points in the
lattice, is described.

\section{   Expansion in the High Temperature Limit and the
Grassmann Multivariable Integrals}

The grand canonical partition function of any quantum system in the
high temperature limit can be expanded in terms of $\beta$ as

\vspace {-0.5cm}

\begin{eqnarray}
 {\cal Z}(\beta; \mu) &=& {\rm Tr}( e^{- \beta {\bf K}})
  \nonumber  \\
&= &   \sum_{n=0}^\infty \; \frac{ (-1)^n}{n!} \,{\rm Tr}[{\bf K}^n]
\, \beta^n,
              \label{1} 
\end{eqnarray}

\noindent where {\bf K} is given by

\begin{equation}
 {\bf K} = {\bf H} - \mu {\bf N}, \label{2}
\end{equation}

\noindent {\bf H} is the hamiltonian of the system, $\mu$ is the
chemical potential and {\bf N} is the total number of particles
operator. 

\vspace{0.5cm}

The  fermionic  creation (${\bf a}_i^\dagger$) and 
destruction (${\bf a}_j$) operators can be
mapped into generators of the Grassmann algebra 
$\{\bar{\eta}_i , \eta_j \}$ as follows \cite{itzykson,swanson,das}:

\begin{equation}
 {\bf a}_i^\dagger \rightarrow \bar{\eta}_i 
\hspace{1cm} {\rm and}
\hspace{1cm} {\bf a}_j \rightarrow \frac{\partial}{ \partial
\bar{\eta}_j}, \label{3}
\end{equation}

\noindent where $i,j= 1,2, \cdots, {\cal N}$. 
The generators of this Grassmann algebra of dimension 
$2^{2{\cal N}}$,  written explicitly as 
$\{ \bar{\eta}_1, \cdots , 
\bar{\eta}_{\cal N}; \eta_1, \cdots, \eta_{\cal N} \}$,  
satisfy the following anti--commutation relations:

\begin {equation}
\{{\eta}_i , \eta_j \} = 0, \hspace {0.5cm} \{ \bar{\eta}_i ,\bar\eta_j
\} =0 \hspace {0.5cm} {\rm and} \hspace {0.5cm} \{ \bar{\eta}_i , \eta_j
\}=0.  \label{4}
\end{equation}

The trace of any normal-ordered fermionic operator {\bf O} 
is\cite{itzykson}

\begin{equation}
 {\rm Tr} [{\bf O}] = \int \prod_{i=1}^{\cal N} \, d\eta_i
d\bar{\eta}_i \; {\cal O}^{\normal}(\bar{\eta}, \eta) \; \; 
e^{\;\sum\limits_{j=1}^{\cal N} 2\bar{\eta}_j \eta_j},
      \label{5}
\end{equation}

\noindent where we use the shorthand notation: $ \bar{\eta} \equiv
\{\bar{\eta}_1, \cdots, \bar{\eta}_{\cal N}\}$ and $\eta\equiv
\{\eta_1, \cdots,\eta_{\cal N}\}$, and ${\cal O}^{\normal}(\bar{\eta},
\eta)$ is the kernel of the fermionic operator {\bf O} in the normal
order. (By ``normal ordered operator'' we mean an operator in which
all destruction operators are to placed to the right of all creation
operators.) Naively, it can be said that the grassmannian function
${\cal O}^{\normal}(\bar{\eta}, \eta)$ is obtained by replacing ${\bf
a}_i^\dagger \rightarrow \bar{\eta}_i $ and ${\bf a}_i \rightarrow
\eta_i$ in operator {\bf O} \cite{phys_A,itzykson}.

Let us consider from now on the case where the creation and
destruction operators are caracterized by the indices $(\vec{\ell};
\sigma)$, where $\vec{\ell}$ is a $d$-dimensional lattice vector
($d=1,2, 3, \cdots$) and $\sigma$ is the spin component. The
components of vector $\vec{\ell}$ need not to be
orthogonal. This
lattice vector could equally represent either the space vector $\vec{\bf
x}$ or the momentum vector $\vec{\bf k}$.
If the fermionic operator {\bf O} is a product of $n$ 
normal ordered fermionic operators {\bf Q}, we have \cite{phys_A}

\begin{eqnarray}{\rm Tr} [{\bf Q}^n] &=& \int \prod_{\vec{\ell}}
\prod_{\sigma = \pm 1}
\prod_{\alpha=0}^{n-1} \, d\eta_\sigma(\vec{\ell}; \alpha) 
d\bar{\eta}_\sigma(\vec{\ell}; \alpha) \;
e^{\sum\limits_{\vec{\ell}} \sum\limits_{\nu=0}^{n-1}
\bar{\eta}_\sigma(\vec{\ell}; \nu) [\eta_\sigma(\vec{\ell}; \nu) 
- \eta_\sigma(\vec{\ell}; \nu+1)]}
\hspace {0.3cm} \times \nonumber \\
& & \hspace{-2cm} \times 
{\cal Q}^{\normal}(\bar{\eta}_\sigma(\vec{\ell}; 0),
\eta_\sigma(\vec{\ell}; 0)) \, 
{\cal Q}^{\normal}(\bar{\eta}_\sigma(\vec{\ell}; 1),
\eta_\sigma(\vec{\ell}; 1))
\times  \cdots \times
{\cal Q}^{\normal}
(\bar{\eta}_\sigma(\vec{\ell}; n-1), \eta_\sigma(\vec{\ell}; n-1)).
\nonumber \\
 &   &         \label{6}
\end{eqnarray}

\noindent where we define
$\sigma = \uparrow\equiv
+1, \sigma = \downarrow\equiv -1$. The Grassmann variables
in eq.(\ref{6}) satisfy the boundary conditions  

\begin{equation}
\eta_\sigma(\vec{\ell}; n) = - \eta_\sigma(\vec{\ell};0) 
\hspace {0.5cm} {\rm and}  \hspace {0.5cm}
\eta_\sigma(\vec{\ell}; \nu) = 0,
 \hspace {0.7cm }{\rm for}  \;\;\; \nu> n,
    \label{7}
\end{equation}

\noindent with $\sigma = \pm 1$ and $\vec{\ell}$ stands for any 
vector on the lattice. Eq.(\ref{6}) is still valid for a 
product of $n$ ordered ordered operators, not necessarily equal.

Relation (\ref{6}) is used to write the terms of the expansion of the
grand canonical partition function in the high temperature limit as
multivariable Grassmann integrals.  
For a $d$-dimensional fermionic model, the coefficients of the
expansion ${\cal Z}(\beta, \mu)$ in eq.(\ref{1}) become  

\begin{eqnarray}
 {\rm Tr} [{\bf K}^n] &= & \int \prod_{\vec{\ell}}
\prod_{\sigma= \pm 1} \prod_{\alpha=0}^{n-1} \, d\eta_\sigma (\vec{\ell}; \alpha)
 d\bar{\eta}_\sigma(\vec{\ell}; \alpha) \; e^{\sum\limits_{\vec{\ell}}
\sum\limits_{\sigma= \pm 1} \sum\limits_{\nu=0}^{n-1}
\bar{\eta}_\sigma(\vec{\ell}; \nu) [\eta_\sigma (\vec{\ell}; \nu) -
\eta_\sigma (\vec{\ell}; {\nu+1})]} \hspace {0.3cm} \times  \nonumber \\
& &
\hspace {-1.5cm}  \times {\cal K}^{\normal}(\bar{\eta}_\sigma(\vec{\ell};
0),
\eta_\sigma( \vec{\ell}; 0)) \, {\cal
K}^{\normal}(\bar{\eta}_\sigma(\vec{\ell}; 1), 
\eta_\sigma ( \vec{\ell}; 1))\times \cdots \times
 {\cal K}^{\normal}
(\bar{\eta}_\sigma( \vec{\ell}; {n-1}), \eta_\sigma( \vec{\ell}; n-1)),  
  \nonumber\\
  & &   \label{8}
\end{eqnarray}

\noindent  the boundary conditions (\ref{7})  still hold for
the generators $\eta_\sigma(\vec{\ell}; \nu)$.

It is much easier to handle generators with one index, then 
we  map the generators $\eta_\sigma (\vec{\ell}, \nu)$ and 
$\bar{\eta}_\sigma (\vec{\ell}, \nu)$ into single-indexed
 anti-commuting variables. The sum in the exponential on
the r.h.s. of eq.(\ref{8}) can be  written as  

\begin{eqnarray}
\sum_{\vec{\ell}} \sum_{\sigma= \pm 1}
 & & \hspace{-0.9cm} \sum_{\nu=0}^{n-1}
\bar{\eta}_\sigma( \vec{\ell}; \nu) [\eta_\sigma ( \vec{\ell}; \nu) -
\eta_\sigma (\vec{\ell}; {\nu+1})]= \nonumber \\
& \equiv &  \sum_{I, J=1}^{2nN^d}
\bar{\eta}_I \;A_{ I J}\; \eta_J. \label{11}
\end{eqnarray}

\noindent Note that the argument of the exponential on the r.h.s. of
eq.(\ref{8}) is diagonal in the indices $\vec{\ell}$ and $\sigma$.
Having the components of the column vector $\eta_J$ (or the line
vector $\bar{\eta}$) grouped according to the values of $\sigma$, and
then each subset ordered according to $\nu$, and finally each
subsubset ordered according to $ \vec{\ell}$, the matrix {\bf A} will
have the block-structure

\begin{equation}
 {\bf A} = \pmatrix { {\bf A}^{\uparrow \uparrow} & \0barra \cr
 & &   \cr
   \0barra & {\bf A}^{\downarrow \downarrow} \cr },  \label{12}
\end{equation}

\noindent whose entries are matrices of dimension
 $nN^d\times nN^d$ and $d$ is the dimension of
the vector $\vec{\ell}$. The
indices $I,J$ are such that $I,J = 1, 2, \cdots, 2nN^d$,
where $N^d$ is the number of points in the lattice.  
The matrices ${\bf A}^{\uparrow \uparrow}$ and ${\bf A}^{\downarrow
\downarrow}$ are identical. Taking into account the anti-periodic
condition in temperature (\ref{7}) in eq.(\ref{11}), the matrices
${\bf A}^{\sigma \sigma}$, where $\sigma=\uparrow, \downarrow$, are
found to have the following block-structure :

\begin{equation}
 {\bf A}^{\uparrow \uparrow} = {\bf A}^{\downarrow \downarrow} =
\pmatrix { \1barra_{N^d\times N^d} & - \1barra_{N^d\times N^d} &
\0barra_{N^d\times N^d} & \cdots & \0barra_{N^d\times N^d}
 \cr & & & & \cr
%
%
\0barra_{N^d\times N^d} & \1barra_{N^d\times N^d}  &
- \1barra_{N^d\times N^d} & \cdots & \0barra_{N^d\times N^d}  \cr
%
%
& & & & \cr \vdots & & & & \vdots \cr
%
%
 \1barra_{N^d\times N^d} & \0barra_{N^d\times N^d} &
\0barra_{N^d\times N^d} & \cdots & \1barra_{N^d\times N^d} \cr}.
                     \label{13}
\end{equation}

\noindent The symbols $\1barra_{N^d\times N^d}$ and
$\0barra_{N^d\times N^d}$ stand for the identity and null matrices of
dimension ${N^d\times N^d}$, respectively.
Any lattice vector $\vec{\ell}$ can be written as
\begin{equation}
\vec{\ell} = \ell_1 \vec{\bf u}_1 + \ell_2 \vec{\bf u}_2+ 
\cdots + \ell_d \vec{\bf u}_d,
\end{equation}

\noindent where $\ell_i = 1, 2 , \cdots, N$ ($i= 1, 2, \cdots, d$),
and $N$ is the number of points in the lattice in the direction of
the $d$-dimensional basis vector
$\vec{\bf u}_i$. A particular basis for which ${\bf
A}^{\sigma \sigma}$ has the block-form shown in (\ref{13}), yields the
following mapping

\begin{equation}
 \eta_\sigma ( \vec{\ell}; \nu) \longrightarrow
\eta_{[{(1-\sigma)\over 2} n + \nu]N^d + 
\ell_1 + (\ell_2 - 1)N + \cdots + (\ell_d - 1)N^{(d-1)}}
 \hspace {4pt}. \label{10}
\end{equation} 

\noindent The generators $\bar\eta_\sigma ( \vec{\ell}; \nu)$ have an
analogous mapping.  With the newly indexed generators, the expression of 
${\rm Tr}[{\bf K}^n]$ (eq.(\ref{8})) becomes,  

\begin{eqnarray}
 {\rm Tr}[{\bf K}^n] &=& \int \prod_{I=1}^{2nN^d} \, d\eta_I
d\bar{\eta}_I \;\; e^{\sum\limits_{I,J= 1}^{2nN^d} \bar{\eta}_I\; A_{I
J} \; \eta_J} \times \nonumber \\
%
%
 & &\hspace {-1cm} \times
{\cal K}^{\normal} (\bar{\eta}, \eta; \nu=0)\; {\cal K}^{\normal}
(\bar{\eta}, \eta; \nu=1) \; \cdots \; {\cal K}^{\normal} (\bar{\eta},
\eta; \nu=n-1).   \label{14}
\end{eqnarray}

Note that expression (\ref{14}), up to the constant $-\frac{1}{n!}$,
 is the coefficient at order
$\beta^n$ of the expansion in the high-temperature limit of the grand
canonical partition function for any  self-interacting
fermionic model. The specific model to be studied is represented by
the grassmannian function ${\cal K}^{\normal}$, but
the matrix {\bf A} is the same for
all  fermionic models. 
Once the sub-matrices ${\bf A}^{\uparrow \downarrow}$ and ${\bf
A}^{\downarrow \uparrow}$ are null, the multivariable integral (\ref{14})
is equal to the product of the contributions coming from the sectors:
$ \sigma \sigma= \uparrow \uparrow$ and $ \sigma \sigma= \downarrow
\downarrow$ separately.  
The Grassmann functions ${\cal K}^{\normal}$ are polynomials in the
generators of the algebra. Therefore, the r.h.s. of eq.(\ref{14}) are
moments of the multivariable Grassmann gaussian integrals. In reference
\cite{BJP} it is shown  that these integrals can be written as co-factors
of the
matrix {\bf A}.  
  
The integrals in eq.(\ref{14}), for  sector
 $\sigma\sigma = \uparrow \uparrow$, have the form: 

\begin{equation}
M(L,K) = \int \prod_{i = 1}^{\rm nN^d} d\eta_i d\bar\eta_i
 \;
\bar\eta_{l_1} \eta_{k_1} \cdots \bar\eta_{l_m} \eta_{k_m} 
\hspace {3pt}
 e^{ \sum\limits_{ i, j = 1}^{\rm nN^d} \bar\eta_i A_{i j}^{\uparrow
\uparrow}
 \eta_j},   \label{15}
\end{equation}

\noindent with $ L= \{ l_1, \cdots, l_m\}$ and $K= \{k_1, \cdots,
k_m\}$. The products $\bar{\eta}\eta$ are ordered in such a way that
$ l_1< l_2< \cdots< l_m$ and $ k_1< k_2< \cdots< k_m$. 
From reference \cite{BJP},
the result of this type of integrals is equal to:  

\begin{equation}
 M(L,K) = (-1)^{(l_1 + l_2+ \cdots+ \l_m) + (k_1+ k_2+ \cdots+ k_m)}
A(L,K), \label{16}
\end{equation}

\noindent where $A(L,K)$ is the determinant of the matrix obtained
from matrix ${\bf A}^{\uparrow \uparrow}$  by deleting 
the lines $\{ l_1, \cdots, l_m\}$ and
the columns $\{k_1, \cdots, k_m\}$. $M(L,K)$ is a co-factor 
of matrix   ${\bf A}^{\uparrow \uparrow}$.
The Grassmann  integrals to 
be calculated in sector $\downarrow \downarrow$ are the
same type as eq.(\ref{15}).    
Evaluating determinants of non-diagonal matrices of dimension
$nN^d\times nN^d$ is still a hard task, even if we have restricted
ourselves to multivariable integrals of a fixed sector $\sigma\sigma$.
Calculating such determinants is a suitable task for computers, and it
obviously depends on hardware and software resources. Fixing $n$,
for instance, there is an upper practical limit for $N$, so that the
calculation of determinants is feasible. One possibility for
evaluating eq.(\ref{14}) is that of assigning different values for $N$
and, from the results obtained, trying to extrapolate for an arbitrary
value of $N$. If we are lucky, some recursion expression for
eq.(\ref{14}) for all $N$, could be recognized.

Our approach to calculate the integral (\ref{15}) is, for fixed $n$ and
arbritary $N$, to explore the block-structure of matrices
 ${\bf A}^{\sigma \sigma}$, $\sigma=\uparrow$ and $\sigma=\downarrow$,
diagonalizing it through a similarity transformation  

\begin{equation}
{\bf P}^{-1} {\bf A}^{\sigma\sigma} {\bf P} = {\bf D}, \label{17}
\end{equation}

\noindent where the matrix {\bf D} is,  

\begin{equation}
 {\bf D} = \pmatrix{ \lambda_1 \1barra_{N^d\times N^d} &
\0barra_{N^d\times N^d} & \cdots & \0barra_{N^d\times N^d} \cr
\0barra_{N^d\times N^d} & \lambda_2 \1barra_{N^d\times N^d} & \cdots &
\0barra_{N^d\times N^d} \cr
 \vdots & & & \vdots\cr
\0barra_{N^d\times N^d} & \0barra_{N^d\times N^d} & \cdots & \lambda_n
\1barra_{N^d\times N^d} \cr } ,   \label{18}
\end{equation} 

\vspace{0.5cm}

\noindent $\lambda_i$, $i= 1, 2, \cdots, n$, are the eigenvalues of
matrices ${\bf A}^{\sigma\sigma}$, $\sigma= \uparrow, \downarrow$, and
calculate the co-factors of the matrix {\bf D}.  
The $j^{th}$-column of matrix {\bf P} is the eigenvector of ${\bf
A}^{\sigma\sigma}$ associated to the eigenvalue $\lambda_j$. Each
eigenvalue of matrix ${\bf A}^{\sigma\sigma}$ has degeneracy $N^d$. The
matrices are not hermitian, thus some eigenvalues are complex. In
the following, we will be working on the
$\sigma\sigma=\uparrow\uparrow$ sector; however, the results for the
$\sigma\sigma=\downarrow\downarrow$ sector are analogous since ${\bf
A}^{\uparrow\uparrow} = {\bf A}^{ \downarrow\downarrow}$.

We will apply the following transformation of variables,

\begin{equation}
 \eta^\prime = {\bf P}^{-1} \, \eta \hspace {1cm} {\rm and} 
\hspace {1cm}
\bar{\eta}^\prime = \bar{\eta} \,{\bf P}, \label{19}
\end{equation}

\noindent where $\eta^\prime \equiv \{ \eta_1^\prime ,
\cdots,\eta_{nN^d}^\prime \}$ and $ \bar{\eta}^\prime \equiv 
\{\bar{\eta}_1^\prime, \cdots, \bar{\eta}_{nN^d}^\prime \}$. 
The jacobian of the transformation (\ref{19}) is equal to one.  

Due to the fact that ${\bf A}^{\uparrow\uparrow}$ is
 a block-matrix, the matrix {\bf P} also has

a block structure. This fact implies that transformations (\ref{19})
do not mix up lattice indices.

In a schematic way, the integrals M(L,K) (eq.(\ref{15})) become:  

\begin{equation}
 M(L,K) = \int \prod_{i = 1}^{\rm nN^d} d\eta_i
d\bar\eta_i
 \; (\bar\eta {\bf P}^{-1})_{l_1} ({\bf P}\eta)_{k_1}
\cdots (\bar\eta {\bf P}^{-1})_{l_m} ({\bf P} \eta)_{k_m} 
	\hspace {3pt}
e^{ \sum\limits_{ i, j = 1}^{\rm nN^d} \bar\eta_i
 D_{i j} \eta_j},   \label{20}
\end{equation}

\noindent where $D_{i j}$ are the entries of the 
diagonal matrix {\bf D}.  
The expression $M(L,K)$ fits into the form of eq.(\ref{15}), and hence
corresponds to some co-factor of the diagonalized matrix {\bf D}
(eq.(\ref{16})). It is very simple to calculate these co-factors, and
the matrix {\bf P} is the same for any self-interacting fermionic
model.

\vspace{0.5cm}

In Appendix \ref{ApendiceA} we present the derivation of the
eigenvalues and eigenvectors of matrix {\bf D} for arbitrary values of
$n$ and $N$. From eqs. (\ref{PsAndQs}) and (\ref{PsAndQs2}),
 for arbitrary value of $n$, we have that

\begin{equation}
 p_{\nu \nu^\prime}^{(n)} = {1 \over \sqrt{n}}
e^{{i \pi \over n} (2\nu^\prime + 1)(\nu +1)},   \label{21}
\end{equation}  
\noindent and
\begin{equation}
q_{\nu^\prime \nu}^{(n)} = {1 \over \sqrt{n}}
e^{-{i \pi \over n} (2\nu^\prime + 1)(\nu +1)},  \label{22}
\end{equation}

\noindent with $\nu, \; \nu^\prime = 0, 1, \cdots, n-1$, and  

\begin{equation}
 {\bf P} = \pmatrix{ p_{0 0}^{(n)}\1barra_{N^d\times N^d} &
 \cdots & p_{0, n-1}^{(n)} \1barra_{N^d\times N^d} \cr
 \vdots & & \vdots\cr
p_{n-1, 0}^{(n)}\1barra_{N^d\times N^d} & \cdots & p_{n-1, n-1}^{(n)}
\1barra_{N^d\times N^d} \cr } \label{23}
\end{equation}

\noindent and    

\begin{equation}
 {\bf P}^{-1} = \pmatrix{ q_{0 0}^{(n)}\1barra_{N^d\times N^d} &
 \cdots & q_{0, n-1}^{(n)} \1barra_{N^d\times N^d} \cr
 \vdots &  & \vdots\cr
q_{n-1, 0}^{(n)}\1barra_{N^d\times N^d} &  \cdots & q_{n-1, n-1}^{(n)}
\1barra_{N^d\times N^d} \cr }.      \label{24}
\end{equation}

\vspace{0.5cm}

\noindent The diagonal elements of matrix {\bf D} are:  \par

\vspace{-0.5cm}

\begin{equation}
\lambda_\nu^{(n)} = 1 - e^{{i\pi \over n} (2\nu +1)},
\hskip 1cm \nu = 0, 1, \cdots, n-1.   \label{25}
\end{equation}

\noindent where the eigenvalues are ${\it N^d}$-fold degenerated,
$N^d$ being the number of lattice sites.
 Due to lattice translation symmetry, we should note that the
elements $p_{\nu \nu^\prime}^{(n)}$ and
 $q_{\nu \nu^\prime}^{(n)}$ do not
carry any lattice site index.

This is a general approach, and it can be applied to any
self-interacting  fermionic model. 
The  important point here is  that the relations (\ref{21})-(\ref{25})
 are valid  for any  self-interacting fermionic
 model with space translation symmetry.

\section{Application to Hatsugai-Kohmoto Model}

The calculation of an exactly soluble model is a nice way to test a
new approach. Hatsugai and Kohmoto\cite{HK} proposed a toy model (HK
model) that shares the atomic and band limits of the Hubbard
model\cite{hubbard}. Using the Green function and path integral
approaches, Nogueira and Anda\cite{nogueira} established the
equivalence of this model (with unrestricted hopping) and the Hubbard
model (with infinite-range hopping).

In this section we derive  the grand canonical partition function
of the HK model using the results presented in section 2.
The hamiltonian of the HK model in momentum space is\cite{nogueira}

\begin{eqnarray}
  {\bf H} & = & \sum_{\vec{k}} \sum_{\sigma= \uparrow, \downarrow} \;
\varepsilon(\vec{k}) {\bf n}_\sigma (\vec{k}) + 
U \sum_{\vec{k}} {\bf n}_\uparrow (\vec{k}) {\bf n}_\downarrow (\vec{k})
        \nonumber \\
  &   &   \nonumber\\
    & \equiv & \sum_{\vec{k}} {\bf H} (\vec{k}),   \label{26}
\end{eqnarray}

\noindent where ${\bf n}_\sigma (\vec{k}) \equiv 
{\bf a}_\sigma^\dagger (\vec{k}) {\bf a}_\sigma (\vec{k})$ and 
${\bf a}_\sigma^\dagger (\vec{k})$ (${\bf a}_\sigma (\vec{k})$) is
the creation (destruction)  operator of an electron
with momentum $\vec{k}$ and spin $\sigma$. The function
$\varepsilon (\vec{k}) = -2t \sum_{i=1}^3  \cos k_i$, 
$\vec{k} = ( k_1, k_2, k_3)$, corresponds to the nearest hopping 
of the electrons in the dual-space lattice. $U$ is the strength of the 
repulsion between  electrons with the same momentum $\vec{k}$
but opposite spin conponents.

From eqs. (\ref{1}) and (\ref{26}), the grand canonical partition
function of the HK model is

\begin{eqnarray}
 \cal{Z} (\beta; \mu) & = & Tr \bigl[\, 
\prod_{\vec{k}} e^{-\beta {\bf K}(\vec{k})} \,\bigr] \nonumber  \\
& = & \prod_{\vec{k}} \;\bigl[Tr\!\!_{\atop{\vec{k}}}   \;
 e^{-\beta {\bf K}(\vec{k})} \bigr].   \label{27}
\end{eqnarray}

\noindent We have 

\begin{equation}
{\bf K} (\vec{k}) = 
\sum_{\sigma=\uparrow, \downarrow} \Delta(\vec{k}) \, {\bf n}_\sigma
(\vec{k}) +
U {\bf n}_\uparrow (\vec{k})  {\bf n}_\downarrow (\vec{k}),  \label{28}
\end{equation}

\noindent where we define: 
$\Delta (\vec{k}) \equiv \varepsilon (\vec{k}) - \mu$,
and $\mu$ is the chemical potencial.
In eq.(\ref{27}), the symbol $Tr\!\!_{\atop{\vec{k}}}$ stands for the
trace for a fixed vector $\vec{k}$, whereas $Tr$ represents the trace
for all $\vec{k}$'s.

The high temperature expansion for the grand canonical partition 
function $\cal{Z} (\beta;\mu)$ is

\begin{equation}
Tr\!\!_{\atop{\vec{k}}} \, \bigl[ e^{-\beta {\bf K}(\vec{k})} \bigr] =
\sum_{n=0}^{\infty}  \frac{ (-1)^n}{n!} \, \beta^n \,
Tr\!\!_{\atop{\vec{k}}}  [{\bf K}^n(\vec{k})]. 
        \label{29}
\end{equation}

\noindent Since all the operators on the r.h.s. of
 eq.(\ref{28}) commute, we 
can apply the Newton's multinomial expression to write 
$Tr\!\!_{\atop{\vec{k}}} [{\bf K}^n(\vec{k})]$ as

\begin{equation}
Tr\!\!_{\atop{\vec{k}}}  [{\bf K}^n(\vec{k})] = 
{\sum\limits_{n_1, n_2, n_3 =0}^{n}}
\vspace{-0.1cm}\hspace{-0.45cm}^\prime
\;\; \;\; \frac{n!}{n_1! n_2! n_3!} \;\;
\Delta^{n_1+ n_2} (\vec{k}) U^{n_3} \;
Tr\!\!_{\atop{\vec{k}}} [ {\bf n}_\uparrow^{n_1+n_3} (\vec{k}) \,
{\bf n}_\downarrow^{n_2+n_3} (\vec{k}) ].  \label{30}
\end{equation}

\noindent The symbol $\sum^\prime$ means that the summation indices
satisfy the condition $n_1+n_2+n_3 = n$.

\vspace{0.5cm}

Let $l_1$, $l_2$ and $l_3$ be the integers that determine the lattice 
vector $\vec{k}$. The mapping (\ref{10}) takes the index $\vec{k}$ into
the index $L\equiv l_1 +  (l_2 -1) N +  (l_3 -1) N^2$, where $N$ 
is the number of points in the momentum lattice in each direction.
In the sum on the r.h.s. of eq.(\ref{30}), we calculate the trace
for a fixed $\vec{k}$, which means
that in eq.({6})  we take a single point in the momentum lattice
$(N=1)$. Then,

\begin{equation}
Tr\!\!_{\atop{\vec{k}}} [ {\bf n}_\uparrow^{n_1+n_3} (\vec{k}) \,
{\bf n}_\downarrow^{n_2+n_3} (\vec{k}) ] = 
{\cal I}_{n_1, n_3}^{\uparrow \uparrow} \times
{\cal I}_{n_2, n_3}^{\downarrow \downarrow}, \label{31}
\end{equation}

\noindent where

\vspace{-1cm}

\begin{eqnarray}
{\cal I}_{n_1, n_3}^{\uparrow \uparrow} & \equiv & 
\int \prod_{I=1}^{n} \, d\eta_I (L) d\bar{\eta}_I (L)  \;\;
e^{\sum\limits_{I,J=1}^n  \bar{\eta}_I (L)  A_{IJ}^{\uparrow \uparrow}
\eta_J (L) }
\times   \nonumber \\
%
&  &  \hspace{-2cm} \times
\bar{\eta}_0(L) \eta_0(L) \cdots \bar{\eta}_{n_1 -1}(L) \eta_{n_1 -1}(L)
\;\;  \bar{\eta}_{n_1+n_2}(L) \eta_{n_1+n_2}(L)  \cdots
\bar{\eta}_{n-1}(L) \eta_{n-1}(L),   \label{32}
\end{eqnarray}

\noindent and

\vspace{-1cm}

\begin{eqnarray}
{\cal I}_{n_2, n_3}^{\downarrow \downarrow} & \equiv & 
\int \prod_{J=n+1}^{2n} \, d\eta_J (L) d\bar{\eta}_J (L)  \;\;
e^{\sum\limits_{I,J=n+1}^{2n} 
 \bar{\eta}_I (L)  A_{IJ}^{\downarrow \downarrow} \eta_J (L) }
\times   \nonumber \\
%
&  &  \hspace{-2cm} \times
\bar{\eta}_{n+n_1}(L) \eta_{n+n_1}(L) \cdots 
\bar{\eta}_{n+n_1+n_2 -1}(L) \eta_{n+n_1+n_2 -1}(L)
   \times \nonumber \\
%
&  \times & 
  \bar{\eta}_{n+n_1+n_2}(L) \eta_{n+n_1+n_2}(L) \cdots
\bar{\eta}_{2n-1}(L) \eta_{2n-1}(L).   \label{33}
\end{eqnarray}

\noindent The matrices ${\bf A}^{\sigma \sigma}$, $\sigma= \uparrow,
\downarrow$, are given by eq.({11}) with $N=1$.
According to eqs. (\ref{15}) and (\ref{16}), the presence of
$\bar{\eta}$'s (and $\eta$'s) in the integrand on the r.h.s. of eqs.
(\ref{32}) and (\ref{33}) allows one to evaluate the integrals as the
determinants of matrices obtained after deletion of lines (and
columns) of the matrices ${\bf A}^{\sigma \sigma}$, $\sigma= \uparrow,
\downarrow$. For this particular model, it turns out easier to apply
eq.(\ref{16}) directly, rather than using the similarity
transformation (\ref{17}), since the lattice is unidimensional. We
should mention that for $N=1$, we recover the case of the anharmonic
fermionic oscillator, which has been considered in a previous work
\cite{phys_A}.
Now we discuss the  values of ${\cal I}_{n_1, n_3}^{\uparrow
\uparrow}$ (eq.(\ref{32})), in view of the possible
values of the indices ($ n_1, n_2, n_3)$.

\vspace{0.3cm}

\noindent {\it i)} $n_1 = n$, $ n_2=0$ and $n_3 =0$.

In this case the first $n$ lines and the first $n$ columns of matrix
${\bf A}^{\uparrow \uparrow}$ are deleted; hence,

\begin{equation}
{\cal I}_{n, 0}^{\uparrow \uparrow} = 1.  \label{34}
\end{equation}

\vspace{0.3cm}

\noindent {\it ii)} $n_1 = 0$, $ n_2= n$ and $n_3 =0$.

In this case no lines or columns are deleted in ${\bf A}^{\uparrow
\uparrow}$; so,

\begin{equation}
{\cal I}_{0, 0}^{\uparrow \uparrow} =
\det({\bf A}^{\uparrow \uparrow}) =2.   \label{35}
\end{equation}

\vspace{0.3cm}

\noindent {\it iii)} $n_1 = 0$, $ n_2= 0$ and $n_3 =n$.

This case is equal to case {\it i} and therefore

\begin{equation}
{\cal I}_{0, n}^{\uparrow \uparrow} = 1.  \label{36}
\end{equation}

\vspace{0.3cm}

\noindent {\it iv)} $n_1 \neq 0$, $ n_2 \neq 0$ and $n_3 \neq 0$.

In this case, the first $n_1$ lines and columns are deleted, as well
as the last $n_3$ lines and columns of matrix ${\bf A}^{\uparrow
\uparrow}$. The triangular matrix thus obtained has its determinant
equal to 1, for any value of $n$. Then,

\begin{equation}
{\cal I}_{n_1, n_3}^{\uparrow \uparrow}  = 1.   \label{37}
\end{equation}

\noindent Equivalent results are valid for 
${\cal I}_{n_2, n_3}^{\downarrow \downarrow}$.

From the results (\ref{34})-(\ref{37}) and the equivalent results
for ${\cal I}_{n_2, n_3}^{\downarrow \downarrow}$, we have

\begin{equation}
Tr\!\!_{\atop{\vec{k}}} [ {\bf n}_\uparrow^{n_1+n_3} (\vec{k}) \,
{\bf n}_\downarrow^{n_2+n_3} (\vec{k}) ] = 
( 1+ \delta_{n_1+n_3, 0}) \, ( 1+ \delta_{n_2+n_3, 0}) , \label{38}
\end{equation}

\noindent that substituted in eq.(\ref{30}) gives

\begin{equation}
Tr\!\!_{\atop{\vec{k}}}  [{\bf K}^n(\vec{k})] =  
\bigl[ 2 \Delta (\vec{k}) + U \bigr]^n +
2 \Delta^n (\vec{k}).  \label{39}
\end{equation}

\noindent Returning to eqs. (\ref{27}) and (\ref{29}), we finally get

\begin{equation}
{\cal Z} (\beta; \mu) = \prod_{\vec{k}} 
\bigl[  1+ e^{- \beta (2 \Delta(\vec{k}) + U)}
+ 2 e^{- \beta \Delta (\vec{k}) } \bigr],   \label{40}
\end{equation}

\noindent that gives the same free energy density found in reference
\cite{nogueira}.

\section{Conclusions}

Calculations involving fermionic fields do demand some extra care, in
comparison to the manipulation of bosonic fields. For this reason,
fermionic models are usually bosonized, in a strategy designed to
avoid the ``annoying'' fermionic features. However, moments of
grassmannian multivariable integrals can be easily calculated, as
shown in eqs. (\ref{15}) and (\ref{16}). In this paper we have
presented a new approach, based on the explicit use of Grassmann
algebra properties, to the problem of calculating the coefficients of
the high temperature expansion of the grand canonical partition
function for any $d$-dimensional self-interacting fermionic model ($d=
1, 2, 3, \cdots$). We have explored the results (\ref{15}) and
(\ref{16}) and the possibility of performing the similarity
transformation (\ref{17}) for a system with arbitrary dimension $d$
and arbitrary number of lattice points $N^d$. It is important to point
out that the matrices ${\bf A}^{\sigma \sigma}$ ($\sigma= \uparrow,
\downarrow$) are model-independent; they are solely related to
kinetical aspects of the approach. To simplify the notation, we
considered that the number of points in each direction
of the lattice is the same, but the results derived
are still valid if this is not true. The fact that our results are
analytical allows us to obtain the thermodynamical limit for any
self-interacting fermionic model.

As a simple example of application of the method (that does not
explore all of its features, though), we have considered the
Hatsugai-Kohmoto model, which is diagonal in momentum space and had
been solved by other approaches. We have derived its grand canonical
partition function, and obtained the same free energy density found in
the literature \cite{nogueira}.

The most important features of this method appear when the hamiltonian
has non-commuting terms and, consequently, Newton's multinomial
expansion does not apply. Equations (\ref{15})-(\ref{16}) and the
similarity transformation (\ref{17}) then become the keystone of our
analytical results. That is the case of the Hubbard model
\cite{hubbard}; the grand canonical partition function for the
unidimensional version of this model is known 
in integral form\cite{takahashi}. A
closed expression for this function was obtained by Takahashi
\cite{takahashi}, within certain limits only. By the application of
the approach we have presented here, we are currently calculating the
coefficients of that partition function, up to order $\beta^5$ and for
any value of the parameters of the model, as well for any value of the
chemical potential. These calculations will soon be submitted to
publication.

\section{Acknowledgements}

E.V.C.S. thanks CNPq and O.R.S. thanks CAPES for total 
financial support.  S.M.de S.  and  M.T.T. 
 thank CNPq for  partial  financial support. M.T.T. also
 thanks FINEP and FAPERJ  and I.C.C. and S.M.de S. thank FAPEMIG
 for partial financial support.

\appendix

\section*{Appendix}

\section{Calculation of Eigenvalues and Eigenvectors
of Matrix ${\bf A^{\sigma \sigma}}$} \label{ApendiceA}

\newcommand{\ee}[1]{e^{#1}}
\newcommand{\B}[2]{{\bf B}^{[{#1},{#2}]}}
\newcommand{\bb}[2]{{\bf b}^{[{#1},{#2}]}}
\newcommand{\bGamma}{{\bf \Gamma}}

This appendix is devoted to calculating the eigenvalues and
eigenvectors of the matrix ${\bf A}^{\sigma \sigma}$, defined in
eq.(\ref{13}), as well as determining the matrices ${\bf P}$ and ${\bf
P}^{-1}$ that diagonalize it (see eq.(\ref{17})).

The characteristic equation for ${\bf A}^{\sigma\sigma}$ is
\begin{eqnarray} \label{bigdeterminant}
 & \det & 
\pmatrix { (1-\lambda)\ \1barra_{N^d\times N^d} & - \1barra_{N^d\times N^d}
&
\0barra_{N^d\times N^d} & \cdots & \0barra_{N^d\times N^d} &
\0barra_{N^d\times N^d}
\cr & & & & \cr
\0barra_{N^d\times N^d} & (1-\lambda)\ \1barra_{N^d\times N^d} &
- \1barra_{N^d\times N^d} & \cdots & \0barra_{N^d\times N^d} &
              \0barra_{N^d\times N^d}
\cr
\vdots & \vdots & \vdots &  & \vdots & \vdots \cr
\0barra_{N^d\times N^d} & \0barra_{N^d\times N^d} & 
\0barra_{N^d\times N^d} & \cdots & (1-\lambda)\ \1barra_{N^d\times N^d} &
- \1barra_{N^d\times N^d} &    \cr\cr
 \1barra_{N^d\times N^d} & \0barra_{N^d\times N^d} &
\0barra_{N^d\times N^d} & \cdots &
\0barra_{N^d\times N^d} & (1-\lambda)\ \1barra_{N^d\times N^d} \cr} 
   \nonumber  \\
 &  & \hspace{1cm} =  0.
\end{eqnarray}

\noindent Observe that this matrix (of total dimension $nN^d\times nN^d$),
consists of a $n\times n$ block matrix, each block having dimension 
$N^d\times N^d$. Moreover, these blocks are either null matrices
$\0barra_{N^d\times N^d}$ or proportional to the identity matrix
$\1barra_{N^d\times N^d}$. \vskip 5mm

We will demonstrate a useful property of the determinant of a block
matrix in which all blocks are diagonal, such as the previous matrix. 
Take a block-matrix ${\bf M}$ composed of blocks $\B{i}{j}$, namely

\begin{equation}
{\bf M} = \pmatrix{
\B{1}{1} & \B{1}{2} & \cdots & \B{1}{n} \cr
\B{2}{1} & \B{2}{2} & \cdots & \B{2}{n} \cr
\vdots & \vdots & \ddots & \vdots \cr
\B{n}{1} & \B{n}{2} & \cdots & \B{n}{n}
}
\end{equation}

\noindent where each block $\B{i}{j}$ is diagonal:

\begin{equation} \label{diagonal}
(\B{i}{j})_{\alpha \beta} = \delta_{\alpha \beta} \ \bb{i}{j}_{\alpha},
\end{equation}

\noindent where $\alpha, \beta=1,2,...,N^d$. (No summation over repeated
indices
is implied). We define a ``determinant-like'' matrix function
${\bf F}$ upon the blocks $\B{i}{j}$ as

\begin{equation} \label{Fdefinition}
{\bf F} \equiv \sum_{\theta_1,\theta_2,\dots ,\theta_n = 1}^{n} \
\varepsilon_{\theta_1,\theta_2,\dots ,\theta_n} \
\B{1}{\theta_1} \ \B{2}{\theta_2}  \dots \B{n}{\theta_n},  
\end{equation}

\noindent where $\varepsilon_{\theta_1,\theta_2,\dots ,\theta_n} $ is the 
Levi-Civita symbol in $n$-dimension.
Obviously, ${\bf F}$ and the blocks $\B{i}{j}$ have all the same
dimensions, $N^d \times N^d$. Using eq.(\ref{diagonal}), we have

\begin{equation} \label{Felement}
F_{ab} =
\delta_{a,b}
\sum_{\theta_1,\theta_2,\dots ,\theta_n = 1}^{n} 
\varepsilon_{\theta_1,\theta_2,\dots ,\theta_n}
(\bb{1}{\theta_1}_{a} \ \bb{2}{\theta_2}_{a}  \dots
     \bb{n-1}{\theta_{n-1}}_{a} \ \bb{n}{\theta_n}_{a}).
\end{equation}

\noindent From eq.(\ref{Felement}) and the definition of the determinant,
we obtain
\begin{equation} \label{detFdetGamma}
\det {\bf F} = \det \bGamma_1 \ \det \bGamma_2 \dots
     \det \bGamma_n,  
\end{equation}

\noindent where we have defined $n$ matrices ${\bf \bGamma}_p$ of dimension
$N^d\times N^d$ as

\begin{equation}
(\bGamma_p)_{uv} \equiv \bb{u}{v}_p
\end{equation}

\noindent so that

\begin{equation}
\det \bGamma_p =
\sum_{\omega_1,\omega_2,\dots ,\omega_N = 1}^{N}
\varepsilon_{\omega_1,\omega_2,\dots ,\omega_n} \
(\bb{1}{\omega_1}_{p} \ \bb{2}{\omega_2}_{p}  \dots
     \bb{n-1}{\omega_{n-1}}_{p} \ \bb{n}{\omega_n}_{p})
\end{equation}

\noindent Thus, the evaluation of $\det {\bf F}$ is equivalent to the
evaluation of the determinant of a block matrix $\bGamma$, defined as

\begin{equation}
\bGamma \equiv \pmatrix{
\bGamma_1 & \0barra_{N^d \times N^d} & \cdots & \0barra_{N^d \times N^d} \cr
\0barra_{N^d \times N^d} & \bGamma_2 & \cdots & \0barra_{N^d \times N^d} \cr
\vdots & \vdots & \ddots & \vdots \cr
\0barra_{N^d \times N^d} & \0barra_{N^d \times N^d} & \cdots &
      \bGamma_n &  \cr
}.
\end{equation}

\noindent However, $\bGamma$ and ${\bf M}$ only differ by an even
number of permutations of lines and columns! More specifically, ${\bf
M}$ can be recovered from $\bGamma$ if we reorder the {\em lines} of
the latter according to the pattern
\begin{eqnarray}
(1,2,\dots,nN^d) & \longrightarrow &
     (1,\  N^d+1,\  2N^d+1, \dots,\  (n-1)N^d+1, \nonumber \\
 & & \hskip 1mm 2,\  N^d+2,\  2N^d+2,\  \dots,\ 
 (n-1)N^d+2, \dots \nonumber \\
 & & \dots, N^d-1,\  2N^d-1,\  3N^d-2, \dots ,\  nN^d-1, \nonumber \\
 & & N^d,\  2N^d,\  3N^d, \dots,\  nN^d) 
\end{eqnarray}

\noindent - i.e., the $1^{st}$ line is left untouched, the $2^{nd}$ line is
replaced by the $(N^d+1)^{th}$ line, etc.- and then have the {\em
columns} of the resulting matrix reordered in the same fashion. (The
same result is obtained if we reorder columns before lines.) As the
total number of permutations is even, we have

\begin{equation} \label{good}
\det {\bf M} = \det \bGamma .
\end{equation}
Combining (\ref{good}), (\ref{detFdetGamma}) and (\ref{Fdefinition}),
we finally obtain
\begin{equation}
\det {\bf M} = \det {\bf F} = \det ( 
     \sum_{\theta_1,\theta_2,\dots ,\theta_n = 1}^{n} \
     \varepsilon_{\theta_1,\theta_2,\dots ,\theta_n} \
     \B{1}{\theta_1} \ \B{2}{\theta_2}  \dots \B{n}{\theta_n} )
\end{equation}

\noindent In conclusion, if the matrix ${\bf M}$ is composed of diagonal
blocks
$\B{i}{j}$, the determinant of ${\bf M}$ is equal to the determinant
of the matrix ${\bf F}$, defined as a ``determinant-like'' function
upon the blocks $\B{i}{j}$.

\vskip 5mm

Turning our attention back to eq.(\ref{bigdeterminant}), we  expand
the determinant in terms of ``cofactors'', based on the last ``line''
of blocks:

\begin{equation} 
\det \pmatrix{ \cr (-1)^{1+n} \ \1barra_{N^d \times N^d}
 \ (-\1barra_{N^d \times N^d})^{n-1}
     + (-1)^{n+n} \ (1-\lambda) \ \1barra_{N^d \times N^d}
          \ (1-\lambda)^{n-1} \ \1barra_{N^d \times N^d}^{n-1} \cr \cr }=0
\end{equation}

\noindent which yields the characteristic equation

\begin{equation} \label{ReducedEquation}
(1 + (1-\lambda)^n)^{N^d} = 0.
\end{equation}

\noindent There are $n$ distinct eigenvalues $\lambda_k$, each one with
multiplicity $N^d$, given by

\begin{equation} \label{lambdak}
\lambda_k = 1- e^{i\frac{\pi}{n}(2k+1)},
\end{equation}

\noindent where $k=0,1,2,...,n-1$.
Observe that if $\lambda_k$ is an eigenvalue, so is its complex
conjugate: $\lambda_{k}^{*} = \lambda_{n-k-1}$. Let us denote by
 ${\bf V}_k$ an eigenvector of ${\bf A}^{\sigma\sigma}$ associated to
$\lambda_k$. It has the structure

\begin{equation}
{\bf V}_k=
\pmatrix{ {\bf v}_1 \cr
{\bf v}_2 \cr
\vdots \cr
{\bf v}_n \cr
}
\end{equation}

\noindent where each ${\bf v}_i$, $i=1,2,...,n$ is a $1 \times N^d$ 
matrix. We obtain

\begin{equation}
{\bf v}_i = -(1-\lambda_k)^i \ {\bf \xi} \mbox{, \ where \ }  i=1,\dots,
n-1
\end{equation}

\noindent and $\xi = {\bf v}_n$ is an {\em arbitrary} 
column vector of dimension $N^d$. There are $N^d$ possible linearly
independent choices for $\xi$, corresponding to $N^d$ distinct
eigenvectors associated to the same eigenvalue $\lambda_k$. We
choose them to be

\begin{equation}
\xi_k^{(1)} = \pmatrix{-1 \cr 0 \cr \vdots \cr 0 \cr 0}, \ 
\xi_k^{(2)} = \pmatrix{0 \cr -1 \cr \vdots \cr 0 \cr 0}, \ 
\ \cdots , \
\xi_k^{(N^d-1)} = \pmatrix{0 \cr 0 \cr \vdots \cr -1 \cr 0}, \
\xi_k^{(N^d)} = \pmatrix{0 \cr 0 \cr \vdots \cr 0 \cr -1}, 
\end{equation}
so that each $\xi_k^{(l)}$ corresponds to an eigenvector
${\bf V}_k^{(l)}$, where $l=1,2,...,N^d$, associated to the eigenvalue
$\lambda_k$.

The matrix ${\bf P}$ that diagonalizes ${\bf A}^{\sigma\sigma}$ can be
obtained by concatenating all eigenvectors ${\bf V}_k^{(l)}$ for all
eigenvalues $\lambda_k$, $k=0,1,...,n-1$, up to a normalizing factor
$R$:

\begin{equation}
 {\bf P} = \pmatrix{ p_{0 0}^{(n)}\1barra_{N^d \times N^d} &
 \cdots & p_{0, n-1}^{(n)} \1barra_{N^d \times N^d} \cr
 \vdots & & \vdots\cr
p_{n-1, 0}^{(n)}\1barra_{N^d \times N^d} & \cdots & p_{n-1, n-1}^{(n)}
\1barra_{N^d \times N^d} \cr },
\end{equation}

\noindent where

\begin{equation}
 p_{\nu \nu^\prime}^{(n)} = R \
e^{{i \pi \over n} (2\nu^\prime + 1)(\nu +1)},
\end{equation}

\noindent with $\nu, \; \nu^\prime = 0, 1, \cdots, n-1$.
The matrix

\begin{equation}
 {\bf P}^{-1} = \pmatrix{ q_{0 0}^{(n)}\1barra_{N\times N} &
 \cdots & q_{0, n-1}^{(n)} \1barra_{N\times N} \cr
 \vdots &  & \vdots\cr
q_{n-1, 0}^{(n)}\1barra_{N\times N} &  \cdots & q_{n-1, n-1}^{(n)}
\1barra_{N\times N} \cr },
\end{equation}

\noindent where

\begin{equation}
q_{\nu^\prime \nu}^{(n)} = R^{\prime}\ 
e^{-{i \pi \over n} (2\nu^\prime + 1)(\nu +1)},
\end{equation}

\noindent is the inverse of ${\bf P}$, upon a suitable choice of $R$ and
$R^{\prime}$; i.e,

\begin{equation}
R=R^{\prime}=1/\sqrt{n},
\end{equation}

\noindent so that they satisfy the relation
\begin{equation} \label{identidade}
\sum_{\bar{\nu}=0}^{n-1} p_{\nu_1 \bar{\nu}}^{(n)} \
     q_{\bar{\nu} \nu_2}^{(n)} = \delta_{\nu_1 \nu_2}.
\end{equation}

\noindent Hence,
\begin{eqnarray} 
p_{\nu \nu^\prime}^{(n)} & = & \frac{1}{\sqrt{n}} 
e^{{i \pi \over n} (2\nu^\prime + 1)(\nu +1)} \label{PsAndQs}  \\
q_{\nu^\prime \nu}^{(n)} & = & \frac{1}{\sqrt{n}} 
e^{-{i \pi \over n} (2\nu^\prime + 1)(\nu +1)}.  \label{PsAndQs2}
\end{eqnarray}





\end{document}